%% file: paper.tex
\documentclass[10pt, letterpaper]{IEEEtran}

\usepackage{amsmath,amssymb,amsfonts}
\usepackage{graphicx}
\usepackage{textcomp}
\usepackage{xcolor}
\usepackage{multirow}
\usepackage{hyperref}

\usepackage{listings}
\usepackage{algorithm,algorithmicx,algpseudocode}
\usepackage{subcaption}
\usepackage{comment}
\usepackage[switch]{lineno}

\def\BibTeX{{\rm B\kern-.05em{\sc i\kern-.025em b}\kern-.08em
	T\kern-.1667em\lower.7ex\hbox{E}\kern-.125emX}}

\begin{document}

\title{A Scalable Actor-based Programming System for PGAS Runtimes}

\author{
\IEEEauthorblockN{Sri Raj Paul}\\
\IEEEauthorblockA{\textit{Intel Corporation, Austin, USA} \\
sriraj.paul@intel.com}\\
\and
\IEEEauthorblockN{Akihiro Hayashi, Kun Chen, Vivek Sarkar}\\
\IEEEauthorblockA{\textit{Georgia Institute of Technology, Atlanta, USA } \\
\{ahayashi,kunfz,vsarkar\}@gatech.edu}
}
	

\maketitle

\begin{abstract}
\input{abstract}

\end{abstract}

\begin{IEEEkeywords}
Actors, Communication Aggregation, Conveyors, OpenSHMEM, PGAS, Selectors
\end{IEEEkeywords}

\section{Introduction}
\input{introduction.tex}


\section{Background: Communication in PGAS Applications}\label{sec:motivation}
\input{comm_layer.tex}



\section{Our Approach}\label{sec:design}
\input{design.tex}






\section{Implementation}\label{sec:impl}
\input{implementation.tex}

\section{Evaluation}\label{sec:evaluation}
\input{evaluation.tex}


\section{Related Work}\label{sec:related_work}
\input{related.tex}

\section{Conclusions and Future Work}\label{sec:conclusion}
\input{conclusions.tex}



\vspace{0.2cm}
\noindent Artifact Availability:{\fontsize{8.5}{10} \selectfont \url{https://github.com/srirajpaul/hclib/tree/bale_actor/modules/bale_actor}}
\vspace{-0.2cm}

\bibliographystyle{IEEEtran}
\bibliography{IEEEfull,biblio}


\end{document}

%% file: abstract.tex
The Partitioned Global Address Space (PGAS) model is well suited for executing irregular applications on cluster-based systems, due to its efficient support for short, one-sided messages. However, there are currently two major limitations faced by PGAS applications.  The first relates to {\em scalability} --- despite the availability of APIs that support non-blocking operations in special cases, many PGAS operations on remote locations are synchronous by default, which can lead to long-latency stalls and poor scalability.  The second relates to {\em productivity} --- while it is simpler for the developer to express all communications at a fine-grained granularity that is natural to the application, experiments have shown that such a natural expression results in performance that is 20$\times$ slower than more efficient but less productive 
code that requires manual message aggregation and termination detection. 

Separately, the actor model has been gaining popularity as a productive asynchronous message-passing approach for distributed objects in enterprise and cloud computing platforms, typically implemented in languages such as Erlang, Scala or Rust.  To the best of our knowledge, there has been no past work on using the actor model to deliver both productivity and scalability to PGAS applications on clusters.   

In this paper, we introduce a new programming system for PGAS applications, in which point-to-point remote operations can be expressed as fine-grained asynchronous actor messages. In this approach, the programmer does not need to worry about programming complexities related to message aggregation and termination detection. Our approach can also be viewed as extending the classical Bulk Synchronous Parallelism model with fine-grained asynchronous communications within a phase or superstep.
We believe that our approach offers a desirable point in the productivity-performance space for PGAS applications, with more scalable performance and higher productivity relative to past approaches.  Specifically,
for seven irregular mini-applications from the Bale benchmark suite executed using 2048 cores in the NERSC Cori system, our approach shows geometric mean performance improvements of $\geq 20\times$ relative to standard PGAS versions (UPC and OpenSHMEM) while maintaining comparable productivity to those versions.



%% file: introduction.tex
In today's world, performance is improved mainly by increasing parallelism, thereby motivating the critical need for programming systems\footnote{Following standard practice, we use the term, "programming system", to refer to both compiler and runtime support for a programming model.} that can deliver both productivity and scalability for parallel applications.
The {\em Actor Model}~\cite{actor,actors_agha} is the primary concurrency mechanism~\cite{actor_primary} in languages such as Erlang and Scala, and is also gaining popularity in modern system programming languages such as Rust.
Large-scale cloud applications~\cite{actor_cloud} from companies such as Facebook and Twitter that serve millions of users are based on the actor model.
Actors express communication using ``mailboxes''~\cite{actor_taxonomy}; the term, ``selector''~\cite{shams_selector}, has been used to denote an actor with multiple mailboxes.
The actor runtime maintains a separate logical mailbox for each actor. Any actor or non-actor, can send messages to an actor's mailbox.
An important property of communication in Actors/Selectors is their inherent asynchrony, i.e., 
there are no global constraints on the order in which messages are processed in mailboxes.


While many classical HPC applications focused on dense data structures,  there has been a recent increase in the use of sparse data structures for HPC, including those used in graph algorithms, sparse linear algebra algorithms, and machine learning algorithms~\cite{openshmem_ml}.
The Partitioned Global Address Space (PGAS)  model~\cite{pgas} is well suited to such irregular applications due to its efficient support for short, non-blocking one-sided messages and the convenience 
of a non-uniform global address space abstraction which enables the programmer to implement scalable locality-aware algorithms. Notable PGAS programming systems include Co-array Fortran~\cite{caf}, OpenSHMEM~\cite{openshmem}, and Unified Parallel C (UPC)~\cite{upc}.
Distribution of irregular data structures across multiple PEs gives rise to large numbers of fine-grain communications, which can be expressed succinctly in the PGAS model.

However, 
a key challenge for PGAS applications 
is the need for careful aggregation and coordination of short messages to achieve low overhead, high network utilization, and correct termination logic.
Communication aggregation libraries such as Conveyors~\cite{conveyors} can help address this problem by locally buffering fine-grain communication calls and aggregating them into medium/coarse-grain messages. However, the use of such aggregation libraries places a significant burden on programmer productivity and assumes a high expertise level.

In this paper, we introduce a new programming system for PGAS applications, in which point-to-point remote operations can be expressed as fine-grained asynchronous actor messages. In this approach, the programmer does not need to worry about programming complexities related to message aggregation and termination detection. 
Further, the actor model also supports the desirable goal of migrating  computation to where the data is located, which is beneficial for many irregular applications~\cite{thread_migrate_emu}. 

Our approach can also be viewed as extending the classical Bulk Synchronous Parallelism (BSP) model with fine-grained asynchronous communications within a phase or superstep.
Many current HPC execution models have been influenced by the simplicity and scalability of the BSP model, which consists of "supersteps" separated by barriers executing on homogeneous processors. However, the increasing degree of heterogeneity and performance variability in exascale machines has motivated the need for including asynchronous computations within a superstep so as to reduce the number of barriers performed and the total time spent waiting at barriers.

Specifically, this paper makes the following contributions:
\begin{enumerate}
    \item An extension of the BSP model to a Fine-grained-Asynchronous Bulk-Synchronous Parallelism (FA-BSP) model. 
	\item{A new PGAS programming system which extends the actor/selector model to enable asynchronous communication with automatic message aggregation for scalable performance.}
	\item{Design of an automatic communication termination protocol that transfers the burden of termination detection and related communication bookkeeping from the programmer to the selector runtime.}
	\item{An implementation of our approach as extensions to the Habanero-C/C++ task-parallel library (HClib) and the Conveyors communication aggregation library.}
	\item{Development of a source-to-source translator that translates our lambda-based  API for actors to a more efficient class-based API.}
	\item{Our results show a geometric mean performance improvement of 25.59$\times$ relative to the UPC versions and  19.83$\times$ relative to the OpenSHMEM versions, while using 2048 cores in the NERSC Cori system on seven irregular mini-applications from the Bale suite~\cite{bale, conveyors}}
\end{enumerate}


%% file: comm_layer.tex
In this section, 
we summarize two fundamental messaging patterns in PGAS applications, namely \textit{read} and \textit{update}, as well as the Conveyors library that can be used to aggregate messages.
Since the focus of our work is on scalable parallelism, 
we assume a Single Program Multiple Data (SPMD) model in which
each processing element (PE) starts by executing the same code with a distinct rank, as illustrated in the following code examples.

\subsection{Read Pattern}
In this pattern, each PE sends a request for data from a dynamically identified remote location and then processes the data received in response to the request. An OpenSHMEM version of a program using this pattern is shown in \autoref{lst:ig_shmem}. This program reads values from a distributed array named \texttt{data} and stores the retrieved values in a local array named \texttt{gather} based on global indices stored in a local array named \texttt{index}. The corresponding operation can also be performed in MPI using \texttt{MPI\_Get}.

\begin{lstlisting}[caption={An OpenSHMEM program that reads data from a distributed array.}, label={lst:ig_shmem}, escapechar=|]
for(i = 0; i < n; i++){
	int  col = index[i] / shmem_n_pes();
	int pe  = index[i] % shmem_n_pes();
	gather[i] = shmem_g(data+col, pe);
}
\end{lstlisting}

\subsection{Update Pattern}
In this pattern, each PE  updates a remote location at an address that is computed dynamically.
An OpenSHMEM program that updates remote locations is shown in \autoref{lst:histogram_shmem}.
This program updates a distributed array named \texttt{histo} based on global indices stored in each PE's local \texttt{index}  array using atomic increment, thereby creating a histogram.
The corresponding operation can also be performed in MPI using
\texttt{MPI\_Accumulate} or \texttt{MPI\_Get\_Accumulate} or \texttt{MPI\_Fetch\_and\_op}.

\begin{lstlisting}[caption={An OpenSHMEM program that creates a histogram by updating a distributed array.}, label={lst:histogram_shmem}, escapechar=|]
for(i = 0; i < n; i++) {
	int spot = index[i] / shmem_n_pes();
	int PE = index[i] % shmem_n_pes();
	shmem_atomic_inc(histo+spot, PE);
}
\end{lstlisting}

\subsection{Conveyors}
Conveyors~\cite{conveyors} is a C-based message aggregation library built on top of conventional communication libraries such as SHMEM, MPI, and UPC. It provides the following three basic operations:
\begin{enumerate}
	\item \texttt{convey\_push}: attempts to locally enqueue a message for delivery to a specified PE.
	\item \texttt{convey\_pull}: attempts to fetch a received message from the local buffer.
	\item \texttt{convey\_advance}: enables forward progress of communication by transferring buffers.
\end{enumerate}

It is worth noting that both \texttt{push} and \texttt{pull} operations can fail (return false) due to resource constraints. \texttt{push} can fail due to a lack of available buffer space, and \texttt{pull} can fail due to a lack of an available item. Due to these failures, \texttt{push} and \texttt{pull} operations must always be placed in a loop that ensures that the operations are retried.  Further, \texttt{advance} needs to be called to ensure progress and to also help with termination detection.
These complexities place a significant burden on programmer productivity and assumes a high expertise level.
\autoref{tab:motivation} demonstrates that user-directed message aggregation with Conveyors can achieve much higher performance compared to non-blocking operations in state of the art communication libraries/systems, some of which includes automatic message aggregation~\cite{upc_aggr1,upc_aggr2}.
As a result, we decided to use Conveyors as a lower-level library for automatic message aggregation in our programming system.

\begin{table}
	\centering
	\resizebox{\columnwidth}{!}{
	\begin{tabular}{|l|l|l|l|}
		\hline
		&             & NB &Time \\ \hline
		\multirow{2}{*}{Read} & OpenSHMEM   (cray-shmem 7.7.10)     & N     & 35.5        \\ \cline{2-4}
		& OpenSHMEM NBI  (cray-shmem 7.7.10)           &  Y &4.2        \\ \cline{2-4}
		& UPC (Berkley-UPC 2020.4.0) & N &22.6  \\ \cline{2-4}
		& UPC NBI (Berkley-UPC 2020.4.0) & Y & 19.7 \\ \cline{2-4}
		& MPI3-RMA (OpenMPI 4.0.2)   & Y & 25.8         \\ \cline{2-4}
		& MPI3-RMA (cray-mpich 7.7.10)   & Y & 8.3         \\ \cline{2-4}
		& Charm++ (6.10.1,  gni-crayxc w/ TRAM)  & Y & 21.3         \\ \cline{2-4}
		& Conveyors (2.1 on cray-shmem 7.7.10)   & Y & 2.3         \\ \hline
		\multirow{2}{*}{Update}    & OpenSHMEM NBI (cray-shmem 7.7.10)      &  Y & 4.3         \\ \cline{2-4}
		& UPC  (Berkley-UPC 2020.4.0) & N &  23.9         \\ \cline{2-4}
		& MPI3-RMA (OpenMPI 4.0.2)   & Y &88.9         \\ \cline{2-4}
		& MPI3-RMA (cray-mpich 7.7.10)   & Y &$>$300         \\ \cline{2-4}
		& Charm++ (6.10.1,  gni-crayxc w/ TRAM)  & Y & 9.7         \\ \cline{2-4}
		& Conveyors (2.1 on cray-shmem 7.7.10)  & Y & 0.5         \\ \hline
	\end{tabular}
	}
	\caption{Absolute performance in seconds using best performing variants for Read and Update benchmarks on 2048 PEs (64 nodes with 32 PEs per node) in the Cori supercomputer which performs 2$^{23}$ ($\approx$8 million) reads and updates. Each version is annotated with a non-blocking (NB) specifier.}
	\label{tab:motivation}
\end{table}

%% file: design.tex
\subsection{Fine-grained-Asynchronous Bulk-Synchronous Parallelism (FA-BSP) model}

The classical Bulk-Synchronous Parallelism (BSP)~\cite{bsp} model consists of "supersteps" separated by barriers executing on homogeneous processors.  Each processor only performs local computations and asynchronous communications in a superstep, and the role of the barrier is to ensure that all communications in a superstep have been completed before moving to the next superstep.  However, the increasing degree of heterogeneity and performance variability in modern cluster machines has motivated the need for including asynchronous computations within a superstep so as to reduce the number of barriers performed and the total time spent waiting at barriers.  To that end, we propose extending BSP to a Fine-grained-Asynchronous Bulk-Synchronous Parallelism (FA-BSP) model, as follows.

Our proposal is to realize the FA-BSP model by building on three ideas from past work in an integrated approach.  The first idea is the actor model, which enables distributed asynchronous computations via fine-grained active messages while ensuring that all messages are processed atomically within a single-mailbox actor.  For FA-BSP, we extend classical actors with multiple symmetric mailboxes for scalability, and with automatic termination detection of messages initiated in a superstep.  The second idea is message aggregation, which we believe should be performed automatically to ensure that the FA-BSP model can be supported with performance portability across different systems with different preferences for message sizes at the hardware level due to the overheads involved.  The third idea is to build on an asynchronous tasking runtime within each node, and to extend it with message aggregation and message handling capabilities.

\begin{figure}
	\centering
	\includegraphics[width=0.5\textwidth]{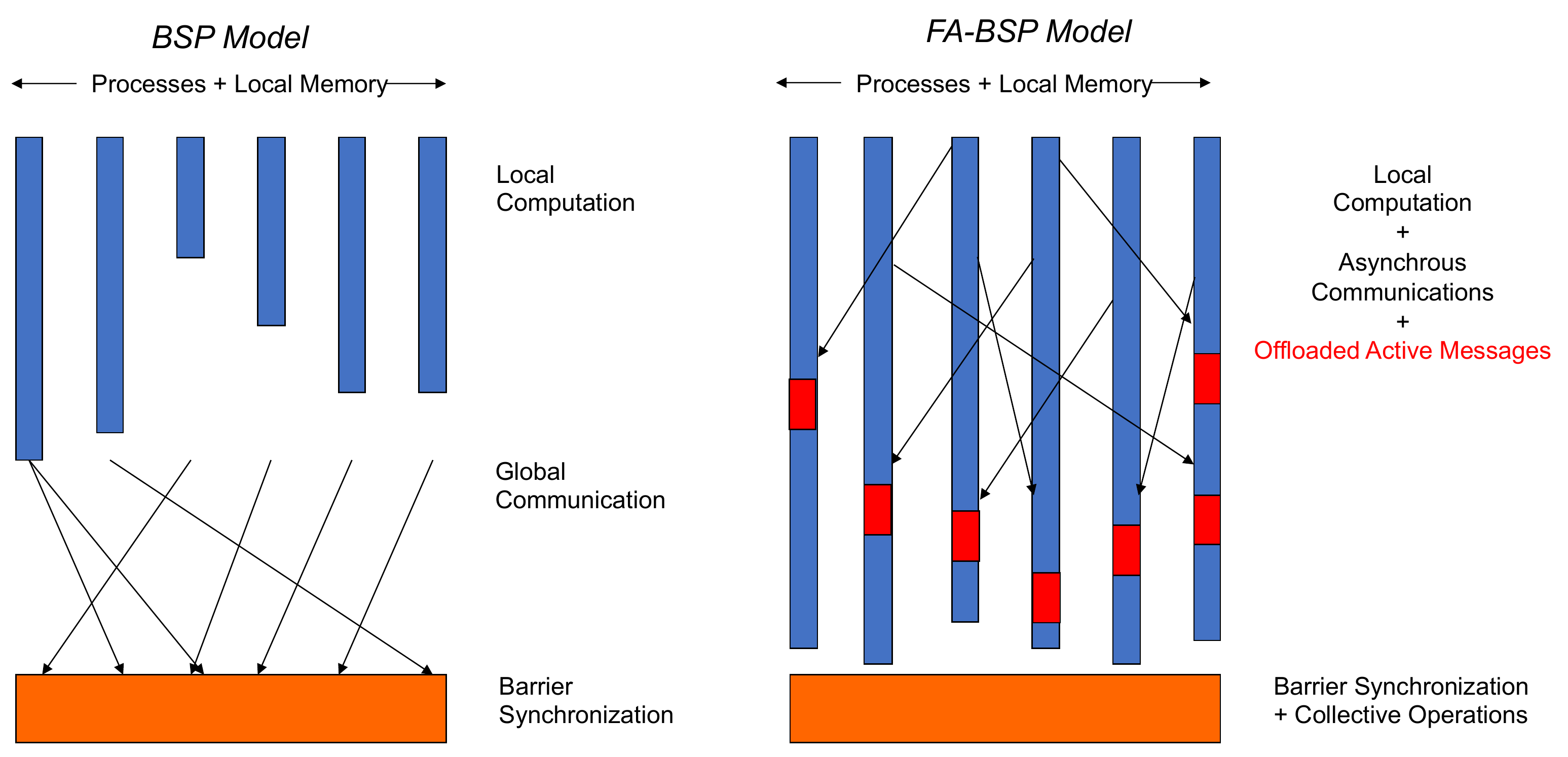}
	\caption{Bulk-Synchronous Parallelism (BSP) vs Fine-grained-Asynchronous Bulk-
Synchronous Parallelism (FA-BSP).  (Graphics adapted from \cite{bsp-slides}.)}
	\label{fig:bsp}
\end{figure}

Figure~\ref{fig:bsp} highlights the key differences between the BSP and FA-BSP models.  We expect the need for fewer barriers in the FA-BSP models since it allows communications to occur within a superstep.

\subsection{High-Level Design of Programming System}\label{subsec:design}
Our primary approach to delivering both productivity and scalability for PGAS applications is by building a programming system based on the actor model that also supports automatic message aggregation and termination detection.
Relative to the Conveyors approach, we would like to remove the burden of the user having to worry about about 1) the lack of available buffer space (\texttt{convey\_push}), 2) the lack of an available item (\texttt{convey\_pull}), and 3) the progress and termination of communications (\texttt{convey\_advance}). We believe that the use of the actor/selector model is well suited for this problem since its programming model productively enables the specification of fine-grained asynchronous messages.  Some key elements of the high-level design are summarized below

\subsubsection{Abstracting \textit{buffers} as \textit{mailboxes}}
We observe that buffer operations can be elevated to actor/selector mailbox operations with much higher productivity.  For example, the \texttt{convey\_push} operation on a buffer can be elevated to an actor/selector \texttt{send} operation, and a \texttt{convey\_pull} operation can be made implicit in an actor/selector's message processing routine, while leaving it to our programming system to handle buffer/item failures and progressing/terminating communications among actors/selectors. More details on how our runtime handles failure scenarios are given in~\autoref{sec:runtime}.


An important design decision for scalability is to treat a collection of mailboxes as a  distributed object so that the mailboxes can be partitioned across PEs, analogous to how memory is partitioned in the PGAS programming model. This partitioned global actor design allows users to access a target actor's mailbox conveniently, 
instead of having to search for the corresponding actor object across multiple nodes as is done in many actor runtime systems.

\subsubsection{Supporting Multiple Mailboxes}
Among the two patterns discussed in \autoref{sec:motivation}, the \textit{read} patterns differs from the \textit{update} pattern in that it involves communication in two directions, namely request and response. Since it is challenging for actors with a single mailbox to implement such synchronization and coordination patterns, this motivates us to instead use a `Selector'~\cite{shams_selector}, which is an actor with multiple mailboxes, as a high-level abstraction. For example, for the \textit{read} operation, users are expected to create two mailboxes (one for Request, the other for Response) and implement the \texttt{Selector}'s message processing functions for the two mailboxes. This partitioned global selector design enables a uniform programming interface across the different communication patterns.

\subsubsection{Progress and Termination}
In general, the Actors/Selectors model provides an \texttt{exit}~\cite{actor_exit} operation to terminate actors/selectors.
While it may seem somewhat natural to expose this operation to users. 
one problem with this termination semantics is that it requires users to  ensure that all messages in the incoming mailbox are processed (or received in some cases) before invoking \texttt{exit}, which adds additional complexities even for the simplest mini-applications such as Histogram~\autoref{lst:histogram_shmem}. To mitigate this burden, we added a relaxed version of \texttt{exit}, which we call \texttt{done}, to enable the runtime do more of the heavy lifting. The semantics of \texttt{done} is that users tell the runtime that the PE on which a specific actor/selector object resides will not send any more messages in the future to a particular mailbox, so the runtime can still keep the corresponding actor/selector alive so it can continue to receive messages and process them. More details on progress and termination can be found \autoref{subsec:termination}.

\subsection{User-facing API}\label{subsec:lambda_api}
\input{api}

\subsection{Termination Graphs}\label{subsec:termination}

\input{termination}

\subsection{Class-based API}\label{subsec:class_api}
\input{class-api}

%% file: api.tex
Based on the discussions in \autoref{subsec:design}, we provide a C/C++ based actor/selector programming framework as shown in~\autoref{lst:actor_selector}. 

\begin{lstlisting}[caption={Actor/Selector Interface with partitioned global mailboxes.}, label={lst:actor_selector}, escapechar=|]
//L: lambda type

class Actor<L> {
  void send(int PE, L msg);
  void done();
};

class Selector<N, L> { // N mailboxes
  void send(int mailbox_id, int PE, L msg);
  void done(int mailbox_id);
};
\end{lstlisting}

\noindent\textbf{Update:} \label{sec:update_pattern}
\autoref{lst:histogram_lambda} shows our version of the histogram benchmark. 
We use C++ lambdas to succinctly describe both the message and its processing routine.
The main program creates an Actor object as a collective operation in Line:~\ref{line:histo_l_obj}, which is used for communication. Then to create the histogram, it finds the target PE in Line:\ref{line:histo_l_pe} and local index within the target in Line:~\ref{line:histo_l_loc} from the global index. Then it sends a message lambda to the target PE's mailbox using the \texttt{send} API. Once the target PE's mailbox gets the message, the actor invokes it, which updates the \texttt{histo} array. Note that the lambda automatically captures the value of \texttt{spot} inside it.  Also, the code for the lambda does not need to be communicated since it is compiled ahead of time and available on nodes.

\begin{lstlisting}[caption={Actor version of the Update benchmark (Histogram) using lambda.}, label={lst:histogram_lambda}, escapechar=|]
Actor h_actor; |\label{line:histo_l_obj}|
for(int i=0; i < n; i++) {
	int spot = index[i] / shmem_n_pes();  |\label{line:histo_l_loc}|
	int possibly_remote_PE = index[i] % shmem_n_pes();  |\label{line:histo_l_pe}|
	h_actor.send(possibly_remote_PE,[=](){histo[spot]+=1;}); |\label{line:histo_send_lambda}|
}
h_actor.done(); |\label{line:histo_l_done}|
\end{lstlisting}

\noindent\textbf{Read:} \label{sec:gather_pattern}
\autoref{lst:ig_lambda} shows how the read pattern can be implemented using our selector-based approach with multiple mailboxes. (Recall that an actor is simply a selector with one mailbox.). 
Here the selector sends the message to the \texttt{Request} mailbox on the target PE on Line:~\ref{line:ig_send_lambda1}. When the message gets processed in target PE, it fetches the required vales in Line:~\ref{line:ig_l_read} and replies with a message to the \texttt{Response} mailbox of the sender PE in Line:~\ref{line:ig_send_lambda2}. In turn, when the reply message gets processed at the sender PE, it writes the values to the \texttt{gather} array at the appropriate index.

Subparts of the application that involve a large number of latency tolerant communication operations can use our API to achieve high throughput, while other parts of the application can continue using other PGAS interfaces as convenient.

\begin{lstlisting}[caption={Selector version of the Read benchmark using the lambda API.}, label={lst:ig_lambda}, escapechar=|]
enum MB{Request, Response};
Selector<2> r_selector; |\label{line:ig_l_obj}|

int sender_PE = shmem_my_pe();
for(int i=0; i < n; i++) {
	int col = index[i] / shmem_n_pes();  |\label{line:ig_l_loc}|
	int possibly_remote_PE = index[i] % shmem_n_pes();  |\label{line:ig_l_pe}|
	r_selector.send(Request, possibly_remote_PE, [=](){  |\label{line:ig_send_lambda1}|
		int ret_val = data[col]; |\label{line:ig_l_read}|
		r_selector.send(Response, sender_PE,  |\label{line:ig_send_lambda2}|
			[=](){  gather[i] = ret_val; }); |\label{line:ig_l_write}|
	});
}
r_selector.done(Request);  |\label{line:ig_l_done}|
\end{lstlisting}

%% file: termination.tex
As mentioned earlier, we provide the \texttt{done} operation as an alternative to terminating actors/selectors by \texttt{exit}~\cite{actor_exit}. This design is based on our observation that 1) sending messages from a partition can be considered as the active part of communication where the user has to invoke \texttt{send} explicitly, but in contrast, 2) receiving messages is the passive part since the arrival of a message is not directly under the user's control. Therefore, we designed the \texttt{done} termination interface to be more associated with sending of messages to a mailbox and leave it to the runtime to keep track and drain all messages sent to it in the future and also in flight.

One may have noticed that, in~\autoref{lst:ig_lambda}, the \texttt{done} operation is performed only for the \texttt{Request} mailbox and not for \texttt{Response} mailbox. This is possible since the \texttt{Response} mailbox depends on the \texttt{Request} mailbox - i.e., a message is only sent from the \texttt{Request} mailbox to the \texttt{Response} mailbox in Line:~\ref{line:ig_send_lambda1}.

Here we introduce the concept of \textit{Termination Graph} to discuss how this is possible. Let us first define that mailbox Y \textit{depends} on mailbox X if a message is sent to mailbox Y in the process function of mailbox X. 
Based on the dependency relation, in general, we can create a directed graph between mailboxes within a selector. We assume the graph is acyclic, which is sufficient for supporting common irregular applications, and 
We assume an imaginary \texttt{Outside} mailbox, which is a virtual mailbox that does not depend on any mailboxes within the selector. A dependency on \texttt{Outside} mailbox implies a message is received from a non-actor/selector or a different actor/selector from the current distributed one.

Given a termination graph, removing an edge, say the one from X to Y, implies no more messages will be sent to mailbox Y in the process function of mailbox X. Therefore the \texttt{done} operation invoked on a mailbox by the user corresponds to removing all incoming edges to that mailbox since the semantics of \texttt{done} means no more messages will be sent to that mailbox. Using this edge deletion notion, termination of a selector can be formulated as follows.

\textit{
Given a graph whose nodes are mailboxes of a selector and edges represent dependency between those mailboxes, termination of the selector corresponds to the removal of all edges from this graph.
}

The user needs to invoke the \texttt{done} operation for those mailboxes that depend on the \texttt{Outside} mailbox. Using the dependency graph, the runtime can find out when to perform \texttt{done} on the dependent mailboxes, as explained in the next paragraph.


\begin{figure}
	\centering
	\includegraphics[width=0.4\textwidth]{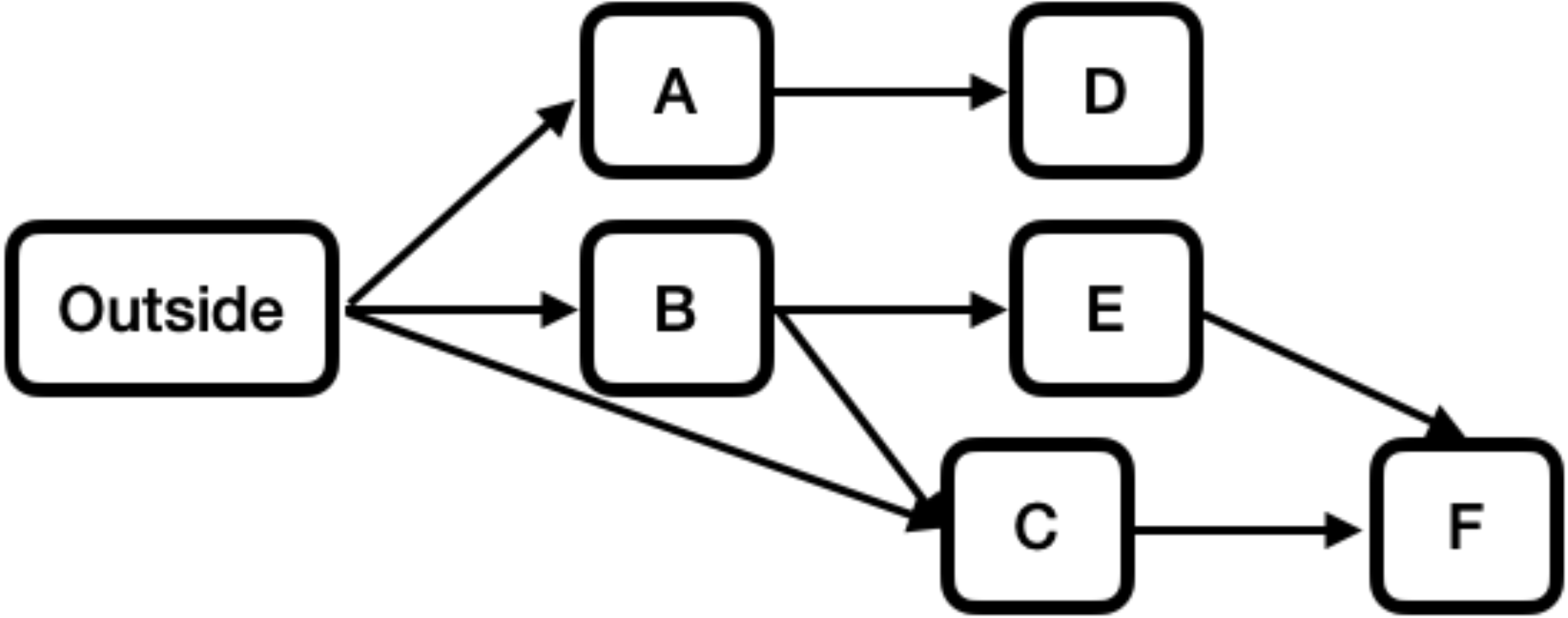}
	\caption{Sample Termination graph with nodes representing mailboxes and edges representing dependencies.}
	\label{fig:termination_graph1}
\end{figure}

~\autoref{fig:termination_graph1} shows a sample mailbox dependency graph in which an arrow from \texttt{A} to \texttt{D} implies that mailbox D depends on mailbox A. For the given figure, the user needs to call \texttt{done} only for the mailboxes A, B and C. The runtime can deduct when to invoke \texttt{done} automatically for the other mailboxes D, E, F. Once the user performs \texttt{done(A)} on a partition, no more sends can be invoked on A from that partition. Still, it can continue to receive and process messages. This implies messages can be sent from the process method any partition of mailbox A to mailbox D. Therefore the runtime needs to ensure \texttt{done(A)} is invoked on all partitions and wait for all messages to be drained from all partitions of mailbox A. At this stage no more messages can be sent to mailbox D since the source of the message is the mailbox of A. Therefore the runtime can now perform \texttt{done} on mailbox D. If a mailbox depends on multiple sources like mailbox F depending on C and E, the runtime waits for the termination of both C and E before invoking \texttt{done(F)}.

For the mini-applications under consideration during evaluation, the only pattern that involved was a linear graph where one mailbox depends on another. Our current implementation uses the linear graph as the default pattern.


%% file: class-api.tex
 While lambdas help with productivity by automatically capturing variables from the environment and enabling the developer to write  routines with in-line message-handling logic instead of separate functions, lambda-based operations can incur additional overhead relative to direct method calls. To avoid this overhead, we also created a class-based version of our APIs that gives the user more control regarding what data needs to be communicated and also allows for automatic translation from the lambda API to the class-based API.


\begin{lstlisting}[caption={Actor/Selector class-based interface with partitioned global mailboxes.}, label={lst:actor_selector_class}, escapechar=|]
class Actor<T> {
  void process(T msg, int PE);
  void send(T msg, int PE);
  void done();
  
  Actor() {
    mailbox[0].process = this->process;
  }
};

class Selector<N,T> { // N mailboxes
  void process_0(T msg, int PE);
  ...
  void process_N_1(T msg, int PE);
  void send(int mailbox_id, T msg, int PE);
  void done(int mailbox_id);
  
  Selector() {
     mailbox[0].process = this->process_0;
     ...
     mailbox[N-1].process = this->process_N_1;
  }
};
\end{lstlisting}

The class-based Actor/Selector interfaces are shown in \autoref{lst:actor_selector_class}. As with the lambda version, the \texttt{Actor.send} API is implemented using \texttt{convey\_push}, and the \texttt{Actor.process} API is implemented using \texttt{convey\_pull} to  process the received message. In the Selector version, we can see there are N message processing routines (\texttt{process\_0} to \texttt{process\_N\_1}), one per mailbox. One important difference that we can notice is that instead of a lambda, the user directly passes the message (\texttt{msg} parameter) that needs to be communicated, and the processing of the message is separately specified using the process routines. A concrete example of how to use the class-based APIs is demonstrated using the Read benchmark in \autoref{lst:ig_lambda_trans}. To put in perspective, if we extend \autoref{tab:motivation} with the actor/selector class-based version, the time taken for Read benchmark is 2.5 sec, and that of Update benchmark is 0.5 sec. Thus we can see that the class-based version can give performance comparable to that of Conveyors and outperform non-blocking communication in the commonly used state-of-the-art communication libraries/systems.

One noticeable difference between the lambda version (\autoref{lst:ig_lambda}) and the class-based version (\autoref{lst:ig_lambda_trans}) is code verbosity and complexity. This is due to class definition boilerplate code and the specification of process methods. To mitigate this issue of verbosity, which inversely affects productivity, we created a source-to-source translator that transforms from a lambda version to a class-based version. The translator also performs optimizations to match the performance of Conveyors. More details can be found in \autoref{subsec:trans}.

%% file: implementation.tex
In this section, we discuss the implementation of the selector runtime prototype created by extending HClib~\cite{grossman2017pulggable}, a C/C++ Asynchronous Many-Task (AMT) Runtime library.
We first discuss our execution model in \autoref{sec:execmodel} and then describe our extensions to the HClib runtime to support our selector runtime in \autoref{sec:runtime}.

\subsection{HClib Asynchronous Many-Task Runtime} \label{sec:hclib}
\input{hclib}

\subsection{Execution Model}\label{sec:execmodel}
\input{execution_model}

\subsection{Selector Runtime}\label{sec:runtime}

\begin{algorithm}
	\caption{Worker loop associated with each mailbox}\label{algo:worker_loop}
\begin{algorithmic}[1]
	
	\While{ buff.isempty() } 
		\State yield() \Comment{yield until message is pushed to buffer} \label{line:yield1}
	\EndWhile
	\State pkt $\gets$ buffer[0]
	\While{ convey\_advance(conv\_obj, is\_done(pkt))} \label{line:advance}
		\For { i $\gets$ 0 to buffer.size-1} \label{line:buff_size}
			\State pkt $\gets$ buffer[i]
			\If { is\_done(pkt)}
				\State break
			\EndIf
			\If {convey\_push(conv\_obj, pkt.data,pkt.rank)} \label{line:push}
				\State break
			\EndIf
		\EndFor
		
		\State buffer.erase(0 to i)
		\While{convey\_pull(conv\_obj, \&data, \&from)} \label{line:pull}
			\State create computation\_task(
			\State \hspace{1em}	process(data, from)\label{line:process}
			\State )
		\EndWhile
		\State yield()
		
	\EndWhile
	\State end\_promise.put(1) \Comment {To signal completion of mailbox} \label{line:put}
\end{algorithmic}
\end{algorithm}

The implementation details presented are based on the class-based interface introduced in \autoref{subsec:class_api}, since our results were obtained by coverting the lambda API to the class-based API using the translator described in \autoref{subsec:trans}.
As mentioned earlier, we hide the low-level details of Conveyors operations from the programmer and incorporate them into our Selector runtime instead. To reiterate, such details include maintaining the progress and the termination of communication as well as handling 1) the lack of available buffer space, and 2) the lack of an available item. This enables users to only stick with the \texttt{send()}, \texttt{done()}, and \texttt{process()} APIs. The implementation details of these APIs are as follows:

\noindent\textbf{Selector.send()}: We map each mailbox to a conveyor object. Each \texttt{send} in a mailbox gets eventually mapped to a \texttt{conveyor\_push}. Note that the \texttt{send} does not directly invoke the \texttt{conveyor\_push} because we want to relieve the computation task on which the application is running from dealing with the failure handling of \texttt{conveyor\_push}. Instead, this API adds a packet with the message and receiver PE's rank to a small local buffer\footnote{This local buffer is different from the Conveyor's internal buffer.} that is based on the Boost Circular Buffer library~\cite{boost_circular}. The packet is later picked up by the communication task associated with the mailbox and is passed into a \texttt{conveyor\_push} operation. Whenever the mailbox's local circular buffer gets filled, the runtime automatically passes control to the communication task, which drains the buffer, thereby allowing us to keep its size fixed.

\noindent\textbf{Selector.done()}: Analogous to \texttt{send}, when \texttt{done} is invoked, we enqueue a special packet to the mailbox that denotes the end of sending messages from the current PE to that mailbox.

\noindent\textbf{Selector.process()}: When the communication task receives a data packet through \texttt{conveyor\_pull}, 
the mailbox's process routine is invoked.



\noindent\textbf{Worker Loop}: The selector runtime creates a conveyor object for each mailbox and processes them separately within its own worker loop, as shown in Algorithm~\ref{algo:worker_loop}. When a mailbox is started, it creates a corresponding conveyor object (\texttt{conv\_obj}) and a communication task that executes the algorithm shown in Algorithm~\ref{algo:worker_loop}. Initially, the communication task waits for data packets in the mailbox's local buffer, which gets added when the user performs a \texttt{send} from the mailbox partition. During this polling for packets from the buffer, the communication task yields control to other tasks, as shown in Line~\ref{line:yield1}. Once the data is added to the buffer, it breaks out of the polling loop and starts to drain elements from the buffer in  Line~\ref{line:buff_size}. It then pushes each element in the buffer to the target PE in Line~\ref{line:push} until push fails. Then it removes all the pushed items from the buffer and starts the pull cycle. It pulls the received data in Line~\ref{line:pull} and creates a computation task, which in turn invokes the mailbox's process method, as shown in Line~\ref{line:process}. As mentioned before, in case there is only one worker that is shared by all the tasks, we invoke the process method directly without the creation of any computation task. Once we come out of the processing of the received data, the task yields so that other communication tasks can share the communication worker.

Once the user invokes \texttt{done}, a special packet is enqueued to the buffer. When this special packet is processed, the \texttt{is\_done} API in Line~\ref{line:advance} returns true, thereby informing the conveyor object to start its termination phase. Once the communication of all remaining items is finished, the \texttt{convey\_advance} API returns false, thereby exiting the work loop. Finally the communication task terminates and signals the completion of the mailbox using a variable of type \texttt{promise} named as \texttt{end\_promise}, as shown in Line~\ref{line:put}. The signaling of the promise schedules a dependent cleanup task which informs all dependent mailboxes, as shown in  \autoref{fig:termination_graph1} about the termination of the current mailbox. This task also manages a counter to find out when all the mailboxes in the selector have performed cleanup, to signal the completion of the selector itself using a \texttt{future} variable associated with the selector. Since the selector runtime is integrated with the HClib runtime, the standard synchronization constructs in AMT runtimes such as \texttt{finish} scope and \texttt{future} can be used by the user to coordinate with the completion of the selector. Other dependent tasks can use the \texttt{future} associated with the selector to wait for its completion. Users can also wait for completion by using a \texttt{finish} scope; for example each of Lines~\ref{line:histo_l_obj}--\ref{line:histo_l_done} in \autoref{lst:histogram_lambda} and ~\ref{line:ig_l_obj}--\ref{line:ig_l_done} in \autoref{lst:ig_lambda} can be enclosed in \texttt{finish} scopes.

\subsection{Source-to-source translation from Lambda-based to Class-based messaging}\label{subsec:trans}
\input{translator.tex}

%% file: hclib.tex
Habanero C/C++ library (HClib)~\cite{grossman2017pulggable} is a lightweight asynchronous many-task (AMT) programming model-based runtime. It uses a lightweight work-stealing scheduler to schedule the tasks.
HClib uses a persistent thread pool called workers, on which tasks are scheduled and load balanced using lock-free concurrent deques. HClib exposes several programming constructs to the user, which in turn helps them to express parallelism easily and efficiently.

A brief summary of the relevant APIs is as follows:
\begin{enumerate}
	\item \texttt{async}: Used to create asynchronous tasks dynamically.
	\item \texttt{finish}: Used for bulk task synchronization. It waits on all tasks spawned (including nested tasks) within the scope of the finish.
	\item \texttt{promise} and \texttt{future}: Used for point-to-point inter-task synchronization in C++11~\cite{c++_future}. A promise is a single-assignment thread-safe container, that is used to write some value and a future is a read-only handle for its value. Waiting on a future causes a task to suspend until the corresponding promise is \textit{satisfied} by putting some value to the promise.
\end{enumerate}

%% file: execution_model.tex
\begin{figure}
	\centering
	\includegraphics[width=0.4\textwidth]{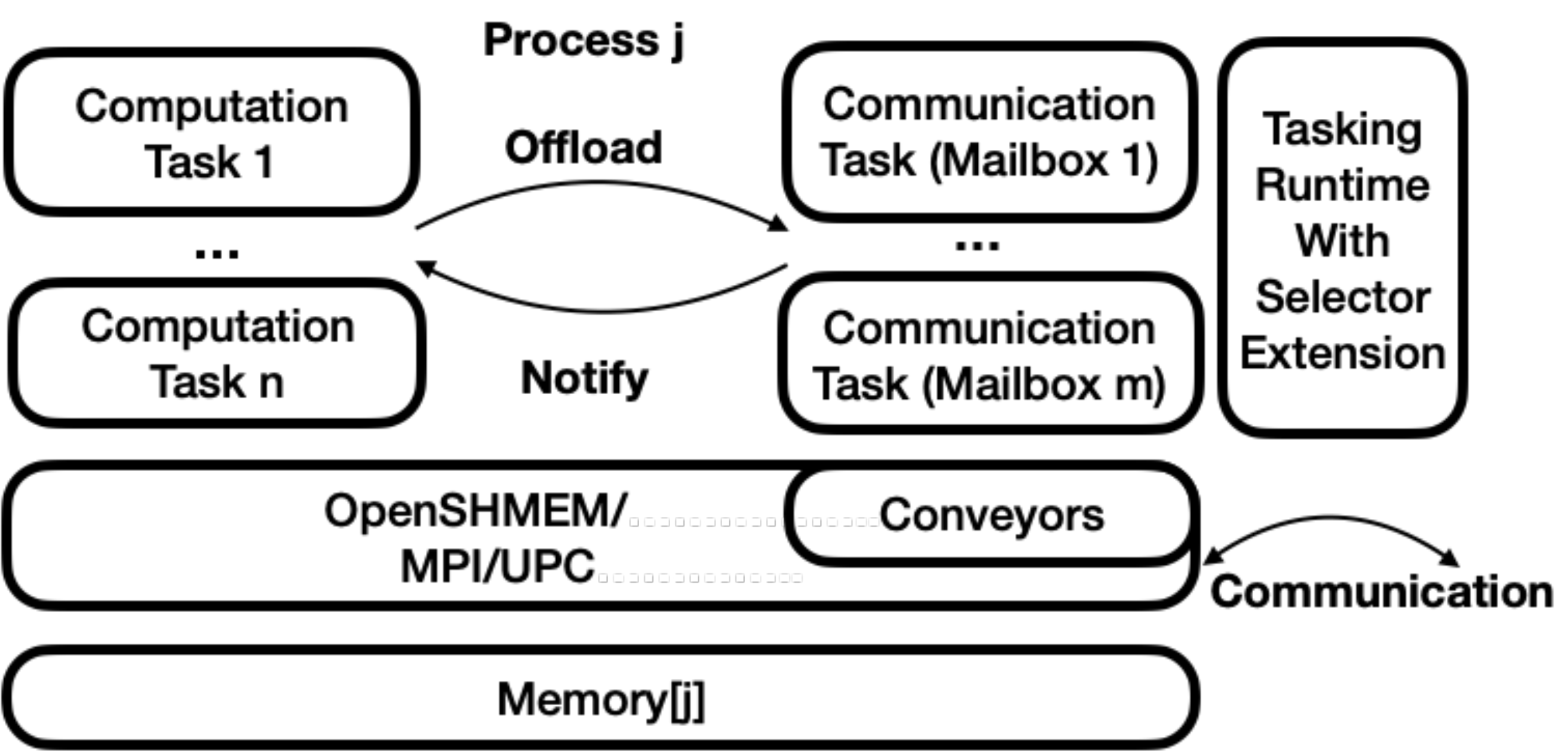}
	\caption{The execution model showing internal structure of tasks and mailboxes within a single PE.}
	\label{fig:execution_model}
\end{figure}

\autoref{fig:execution_model} shows the high level structure of the execution model for our approach from the perspective of PE $j$, shown as process[j], with memory[j] representing that PE’s locally accessible memory. This local memory includes partitions of global distributed data, in accordance with the PGAS model.
Users can create as many tasks as required by the application, which are shown as \textit{Computation Tasks}. For the communication part, each mailbox corresponds to a {\em Communication Task}. All tasks get scheduled for execution on to underlying worker threads. For example, if an application uses a selector with two mailboxes and an actor/selector with one mailbox, it corresponds to three communication tasks --- two for the selector  and one for the actor. All computation and communication tasks are created using the HClib~\cite{grossman2017pulggable} Asynchronous Many-Task (AMT) runtime library.

To enable asynchronous communication, the computation tasks offload all remote accesses on to the communication tasks~\cite{asyncshmem}. When the computation task sends a message, it is first pushed to the communication task associated with the mailbox using a local buffer. Eventually, the communication task uses the conveyors library to perform message aggregation and actual communication. 
Currently we use a single worker thread that multiplexes all the tasks. When a mailbox receives a message, the mailbox's process routine is invoked.

It is worth noting that users are also allowed to directly invoke other communication calls outside the purview of our Selector runtime. For example, the user application can directly invoke the OpenSHMEM barrier or other collectives. 

%% file: translator.tex
While the use of C++ lambda expressions further simplifies writing remote message handlers (\autoref{subsec:lambda_api}), the performance of the lambda-based API is lower than that of the class-based version (\autoref{subsec:class_api}). This motivates us to perform automatic source-to-source translation from the lambda version to the class version to improve productivity without this performance loss.  This kind of translation could be beneficial to other lambda-based libraries as well.

\begin{figure}
	\centering
	\includegraphics[width=0.48\textwidth]{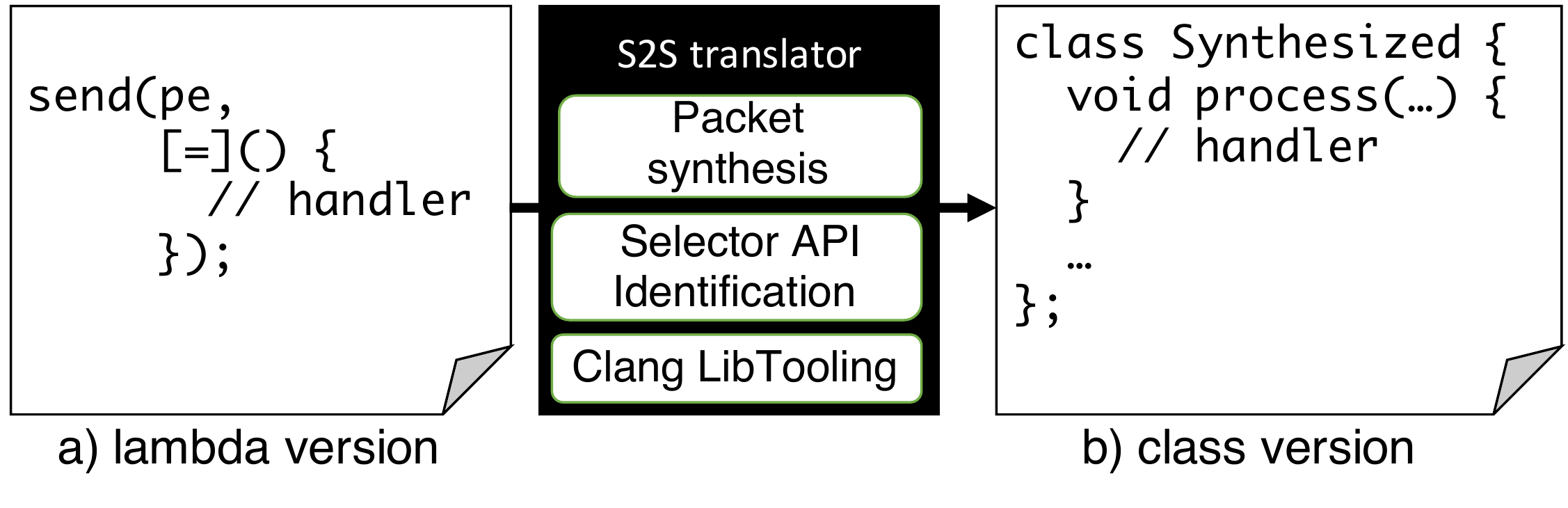}
	\caption{Source-to-source translator from lambda version to class based version.}
	\label{fig:lambda_to_class}
\end{figure}

\autoref{fig:lambda_to_class} illustrates the end-to-end flow for the translation. The translator is a standalone tool built on top of Clang LibTooling. First, it identifies the use of the send API with a lambda expression by utilizing Clang LibTooling's AST traversal APIs. For each lambda, it analyzes captured variables to synthesize a packet structure that is used for the class-based version. Then, it synthesizes a class declaration with a message handler and a packet struct type for actor messages.

\begin{lstlisting}[escapechar=|, caption=An auto-translated version of Index-Gather., label=lst:ig_lambda_trans]
enum MB{Request, Response};
struct packet0 {
  int slot0;
  int slot1;
};
class SynthesizedSelector : public Selector<2, packet0> {
public:
    int64_t *data; int64_t *gather;
    void process0(packet0 pkt, int sender_rank) {
      pkt.slot0 = data[pkt.slot0];
      send(Response, pkt, sender_rank); // pkt.slot1 is unchanged
    }
    void process1(packet0 pkt, int sender_rank) {
      gather[pkt.slot1] = pkt.slot0;
    }
    SynthesizedSelector(int64_t *_ltable, int64_t *_tgt) { ... }
};
SynthesizedSelector r_selector(data, gather);
int sender_PE = shmem_my_pe();
for (int i = 0; i < n ; i++) {
  int col = index[i] / shmem_n_pes();
  int possibly_remote_PE = index[i] % shmem_n_pes();
  packet0 pkt0;
  pkt0.slot0 = col;
  pkt0.slot1 = i;
  r_selector.send(Request, pkt, possibly_remote_PE);
}
\end{lstlisting}

\autoref{lst:ig_lambda_trans} shows an auto-translated version of \autoref{lst:ig_lambda}. Comparing the two programs, one can see that it is feasible to automate this translation, thereby enabling the results in \autoref{sec:evaluation} to be obtained using lambda-based APIs.
All benchmarks were automatically translated from lambda-based versions to class-based versions. 

One interesting challenge that we also overcame was how to synthesize a minimum packet structure for all lambda-based messages. To understand this challenge, let us first revisit the original lambda version in \autoref{lst:ig_lambda}. There is a \texttt{send()} API call with a nested lambda on Line: \autoref{line:ig_send_lambda1}. Notice that, since each invocation of these lambda expressions can happen on a random remote PE, variables captured by a lambda need to be transferred. In this example, the scalar variables \texttt{col} and \texttt{i} are captured by the outer lambda, which is executed on \texttt{possibly\allowbreak \_\allowbreak remote\_\allowbreak PE}. Those variables need to transferred from \texttt{sender\_PE} to \texttt{possibly\allowbreak \_\allowbreak remote\_\allowbreak PE}. Similarly, the scalar variables \texttt{ret\_val} and \texttt{i} are supposed to be transferred back from \texttt{possibly\allowbreak \_\allowbreak remote\_\allowbreak PE} to \texttt{sender\_PE}.
Since the lifetimes of \texttt{col} and \texttt{ret\_val} do not overlap, our translator assigns \texttt{col} and \texttt{ret\_val} to \texttt{slot0} and assigns \texttt{i} exclusively to \texttt{slot1} in \autoref{lst:ig_lambda_trans}.

%% file: evaluation.tex
This section presents the results of an empirical evaluation of our selector runtime system on a multi-node platform to demonstrate its performance and scalability.

\noindent
The goal of our evaluation is twofold:
\begin{enumerate}
\item to demonstrate that our selector-based programming system based on the FA-BSP model can be used to express a range of irregular mini-applications, and
\item to compare the performance of our approach with that of UPC, OpenSHMEM and Conveyors versions of these mini-applications.
\end{enumerate}

\noindent
\textbf{Platform:} We ran the experiments on the Cori supercomputer located at NERSC. In Cori, each node has two sockets, with each socket containing a 16-core Intel Xeon E5-2698 v3 CPU $@$ 2.30GHz (Haswell).
For inter-node connectivity, Cori uses the Cray Aries interconnect with Dragonfly topology that has a global peak bisection bandwidth of 45.0 TB/s. We used cray-shmem 7.7.10, Berkeley UPC 2020.4.0 and GCC 8.3.0 (all available in Cori\footnote{We believe NERSC setup the  modules with the best parameters for Cori.}) to build all the software. Cray-shmem and Berkeley UPC in Cori use Cross-partition memory (XPMEM) technology for cross-process mapping of user-allocated memory within a node that enables load-and-store semantics and native atomics. We used the same  to obtain results in \autoref{tab:motivation}.
We use one worker thread per PE rank for the experiments; since the mini-applications have sufficient parallelism across PE ranks, there was no motivation to use multiple worker threads within a single PE rank.
The Conveyors library was compiled using cray-shmem for our experiments since cray-shmem provided the best performance based on our evaluation in~\autoref{tab:motivation}. Conveyors can also use UPC or MPI as backends, in which case our Selectors library can also be invoked from any UPC or MPI program.

\begin{figure}[t]
	\centering
\begin{subfigure}{.40\textwidth}
	\centering
	\includegraphics[width=1\textwidth]{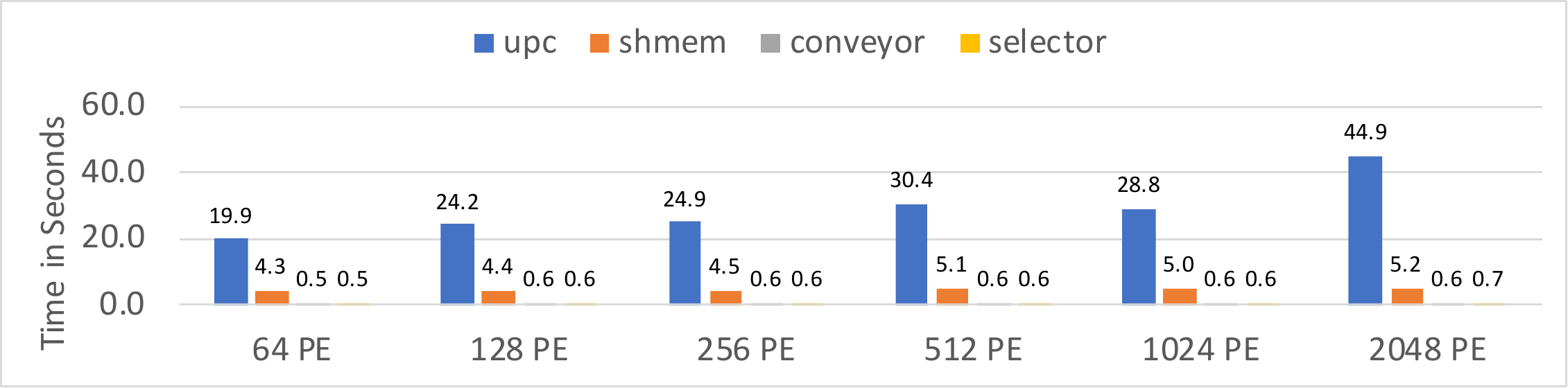}
	\caption{Histogram mini-application with 10,000,000 updates per PE on a distributed table with 1,000 elements/PE.}
	\label{fig:histogram}
\end{subfigure}

\begin{subfigure}{.40\textwidth}
	\centering
	\includegraphics[width=1\textwidth]{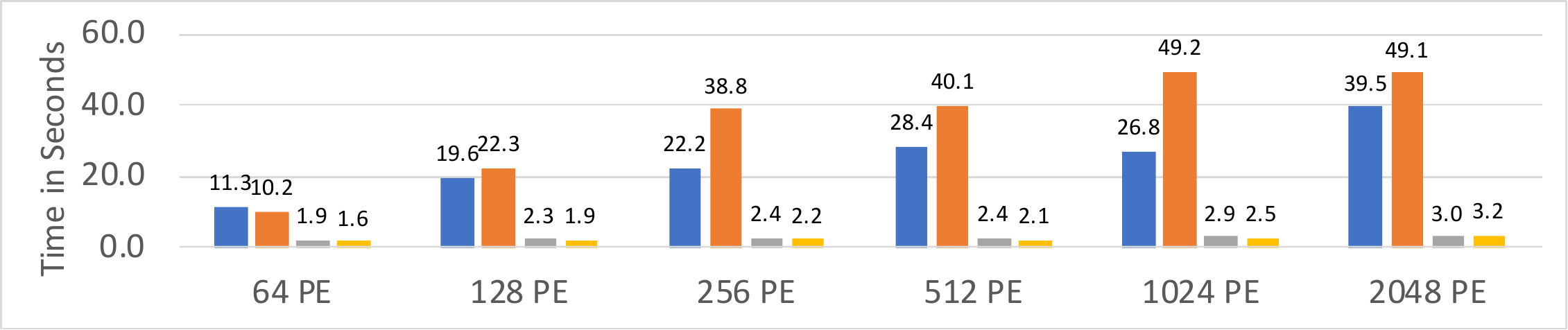}
	\caption{Index-gather mini-application with 10,000,000 reads per PE on a distributed table with 100,000 elements/PE.}
	\label{fig:index-gather}
\end{subfigure}

\begin{subfigure}{.40\textwidth}
	\centering
	\includegraphics[width=1\textwidth]{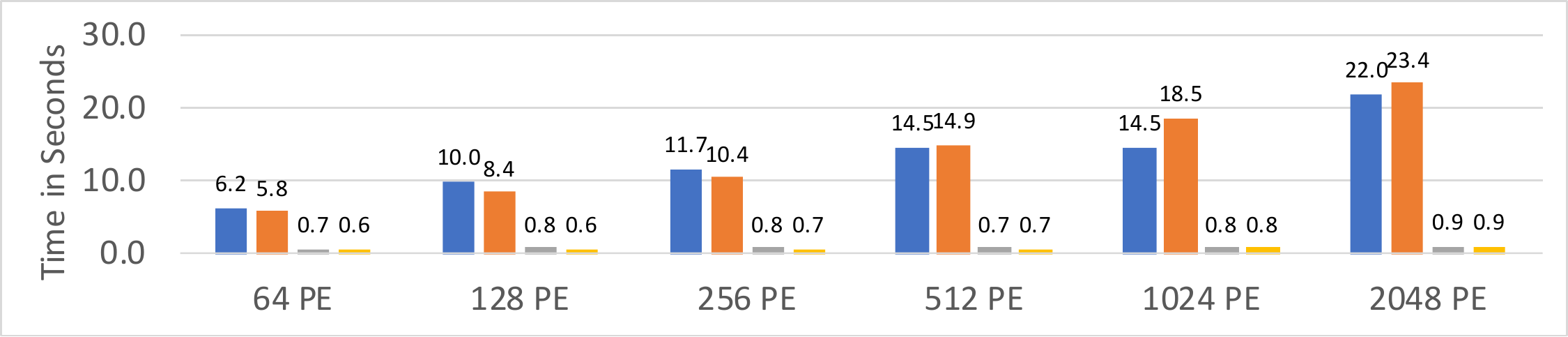}
	\caption{Permute-matrix mini-application with 100,000 rows of the matrix/PE with an average of 10 nonzeros per row.}
	\label{fig:permute-Matrix}
\end{subfigure}

\begin{subfigure}{.40\textwidth}
	\centering
	\includegraphics[width=1\textwidth]{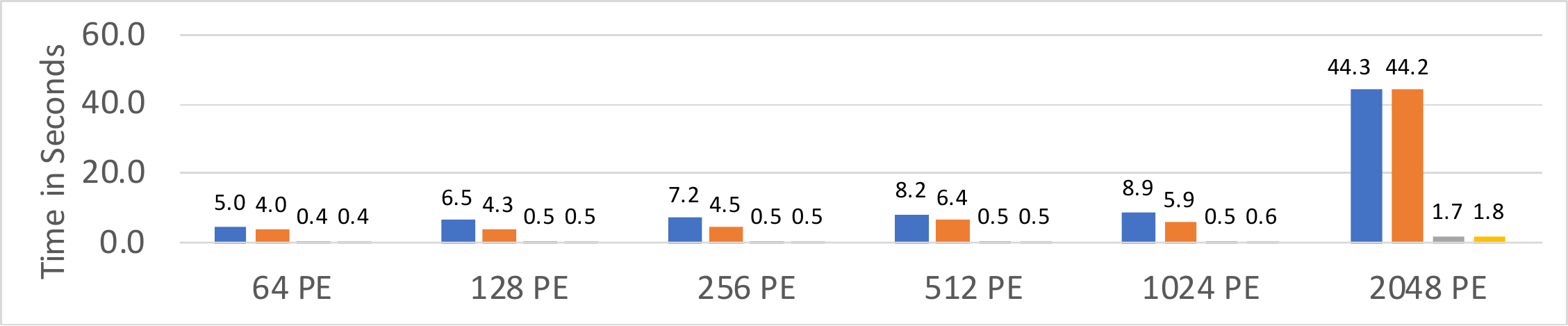}
	\caption{Random-permutation mini-application with 1,000,000 elements per PE.}
	\label{fig:random-permutation}
\end{subfigure}

\begin{subfigure}{.40\textwidth}
	\centering
	\includegraphics[width=1\textwidth]{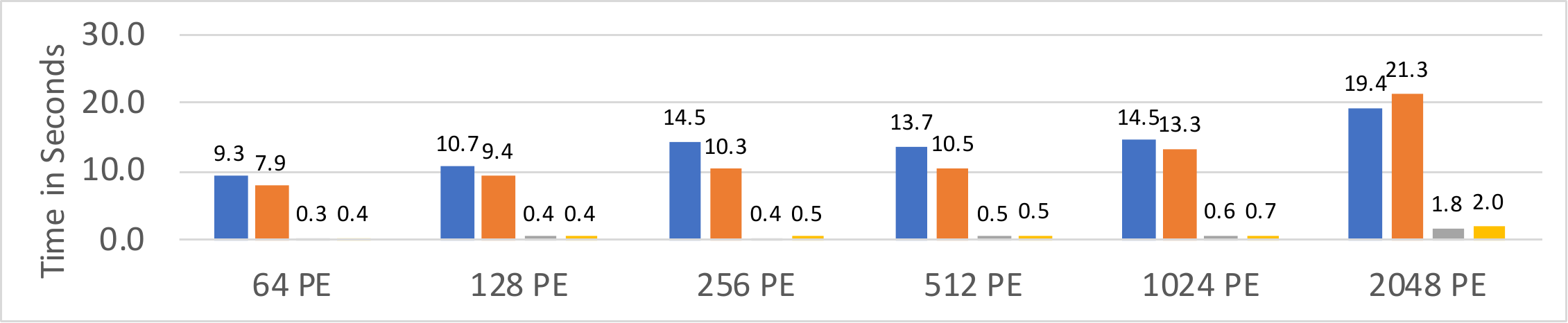}
	\caption{Topological-sort mini-application with 100,000 rows of the matrix/PE with an average of 10 nonzeros per row.}
	\label{fig:topological-sort}
\end{subfigure}

\begin{subfigure}{.40\textwidth}
	\centering
	\includegraphics[width=1\textwidth]{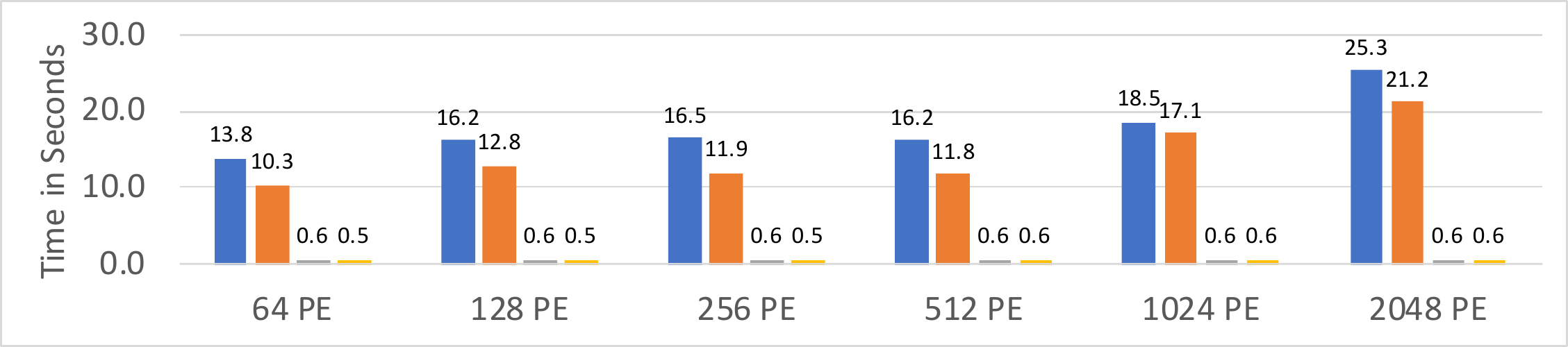}
	\caption{Transpose-matrix mini-application with 100,000 rows of the matrix per PE with an average of 10 nonzeros per row.}
	\label{fig:transpose}
\end{subfigure}

\begin{subfigure}{.40\textwidth}
	\centering
	\includegraphics[width=1\textwidth]{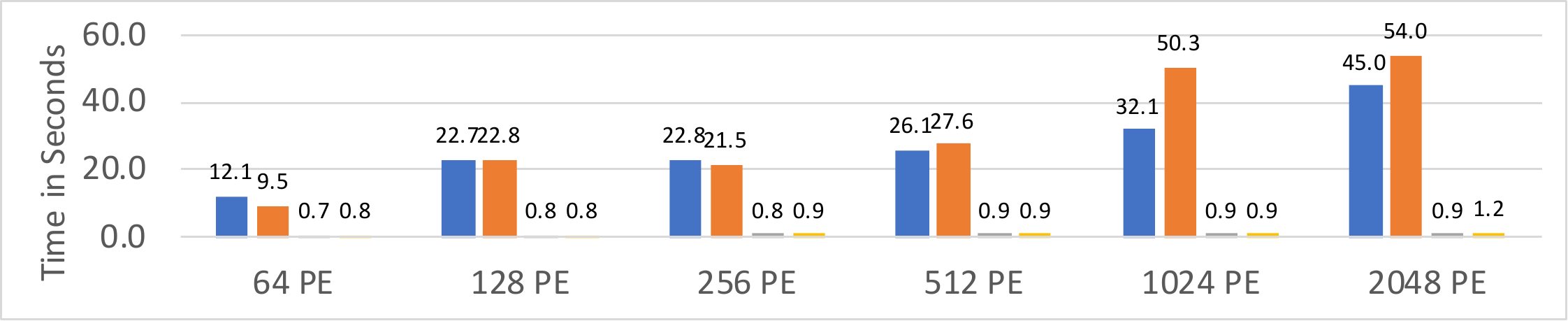}
	\caption{Triangle-counting mini-application with 10,000 rows of the matrix per PE with an average of 35 nonzeros per row.}
	\label{fig:triangle}
\end{subfigure}
\caption{Comparison of execution time of the UPC, OpenSHMEM, conveyor and selector variants (Y-axis: lower is better).}
\label{fig:result}
\end{figure}

\noindent \textbf{Mini-applications:}  We used all seven mini-applications in Bale~\cite{bale,conveyors} that have Conveyors versions for our study.
Bale can be seen as a proxy for key components in an irregular application that involve a large number of irregular point-to-point communication operations.
All scalability results were obtained with weak scaling.

The first mini-application computes a {\em histogram}, using a 
partitioned global array with each PE storing a local table of 1,000 integer elements (similar to the update pattern in \autoref{sec:update_pattern}). Each PE independently performs 10,000,000 atomic increments on this global array.  

The second mini-application performs an {\em index gather} on a partitioned global array.
This mini-application is a straightforward use case of the read pattern mentioned in \autoref{sec:gather_pattern}.

The third mini-application performs {\em permutations} on a distributed sparse matrix. It permutes the rows and columns of the matrix according to two random permutations.

The fourth mini-application solves the {\em random permutation} problem~\cite{random_perm}, which is to generate a random permutation of $1 \ldots m$, assuming that each permutation is equally likely. This mini-application generates a random permutation in parallel using the ``dart-throwing algorithm'' \cite{random_perm}.

The fifth mini-application performs {\em topological sorting} on a distributed sparse matrix. It uses an upper-triangular matrix (with ones on the diagonal) as its input. We then randomly permute the rows and columns of the upper-triangular matrix to obtain a new matrix. Our goal is to find a row and a column permutation such that when these permutations are applied, we can reconstruct an upper triangular matrix.

The sixth mini-application finds the {\em transpose} of a distributed sparse matrix in parallel.

 The seventh mini-application performs {\em triangle counting} in a graph. The graph is represented as an adjacency matrix, which, in turn is stored using a sparse matrix data structure.

\textbf{Experimental variants:}
Each  mini-application was evaluated by comparing the following four versions.  Among these, the UPC and OpenSHMEM versions were obtained from the Bale release~\cite{bale} by replacing calls to libgetput with direct UPC and OpenSHMEM constructs (which resulted in a similar performance to that of the libgetput calls). The Conveyor versions were used unchanged from the Bale release.  All problem sizes are identical to those used in the Bale release with one exception - the average number of nonzeros per row was reduced  from 35 to 10 for all versions of the topological-sorting mini-application to make its execution time more comparable to that of the other mini-applications.
\begin{enumerate}
	\item{\bf UPC:} This version is written using UPC.
	\item{\bf OpenSHMEM:} This version is written using OpenSHMEM.
 	\item{\bf Conveyor:} This version directly invokes the Conveyors APIs, which includes explicit  handling of failure cases and  communication progress.  
	\item{\bf Selector:} This version uses the class-based version of the Selector API introduced in this paper, obtained by automatic translation from the lambda version as described in Section~\ref{subsec:trans}.
\end{enumerate}


In Figures~\ref{fig:result}(a) to \ref{fig:result}(g), the Y-axis shows the execution time in seconds, so smaller is better.  Since we used weak scaling across PEs, ideal weak scaling should show the same time used by a mini-application for all PE counts. From the figures, we can see that the Conveyor versions perform much better than their UPC and OpenSHMEM counterparts.  For the 2048 PE/core case,
the Conveyor versions show a geometric mean performance improvement of 
27.77$\times$ relative to the UPC and 21.52$\times$ relative to the OpenSHMEM versions, across all seven mini-applications.

This justifies our decision to use the Conveyors library for message aggregation in our Selector-based approach.
Overall, we see that the Selector version also performs much better than the UPC/OpenSHMEM versions and close to the Conveyor version.  For the 2048 PE/core case,
the Selector versions show a geometric mean performance improvement of 25.59$\times$ relative to the UPC and 19.83$\times$ relative to the OpenSHMEM  versions, and a geometric mean slowdown of only 1.09$\times$ relative to the Conveyor versions.
These results confirm the performance advantages of our approach, while the productivity advantages can be seen in the simpler programming interface for the Selector versions relative to the Conveyor versions.

\begin{table}[t]
	\centering
	\resizebox{\columnwidth}{!}{
	\begin{tabular}{|l|c|c|c|c|}
		\hline
		& UPC & OpenSHMEM & Conveyor &  Actor/Selector \\ \hline
		Histogram & 18 & 19 &	30 &	21 \\ \hline
        Index-gather & 16 & 17 &	40 &	25 \\ \hline
        Permute-matrix & 37 & 51 &	108 &	78 \\ \hline
        Random-permutation & 41 & 43 &	111	& 99 \\ \hline
        Topological-sort & 72 & 92 &	148 &	130 \\ \hline
        Transpose & 43 & 50 &	83 &	69 \\ \hline
        Triangle-counting &	43 & 49 &	61 &	53 \\ \hline
	\end{tabular}
	}
	\caption{Kernel size of each mini-application in terms of source lines of code (SLOC)}
	\label{tab:productivity}
\end{table}
For completeness, \autoref{tab:productivity} shows the source  lines of code (SLOC) for different versions of the kernel of each mini-application, as measured by the CLOC utility~\cite{cloc}.
The table convincingly shows that moving to the Actor/Selector model results in lower SLOC values relative to the Conveyor model, which in turn demonstrates higher productivity for the Actor/Selector model.

We can also create non-blocking OpenSHMEM operations using Actors, thus enabling message aggregation in those operations. An example using Actors to implement message aggregation enabled OpenSHMEM \texttt{get\_nbi} and \texttt{put\_nbi} operations are given at~\cite{get_nbi_selector, put_nbi_selector}.


%% file: related.tex
The Chare abstraction in Charm++\cite{charm++} has taken inspiration from the Actor model, and is also designed for scalability.  As indicated earlier, the performance of Charm++ is below that of Conveyors (and hence that of our approach) for the workloads studied in this paper.


In the past, there has been much work on optimizing the communication of PGAS programs through communication aggregation.
Avalo \textit{et al}. ~\cite{avlanos} used techniques such as static coalescing and the inspector-executor model to optimize communication in UPC. 
~\cite{akihiro_pgas} introduced new PGAS language-aware LLVM passes to reduce communication overheads. Wesolowski \textit{et al}. ~\cite{tram} introduced the TRAM library that optimizes communication by routing and combining short messages. 
UPC~\cite{upc_aggr1, upc_aggr2} performs automatic message aggregation to improve the performance of fine-grained communication but is unable to achieve performance compared to user-directed message aggregation. We use Conveyors~\cite{conveyors}, which is a modular, portable, efficient, and scalable library as our message aggregation runtime.

%% file: conclusions.tex
This paper proposes a scalable programming system for PGAS runtimes to accelerate irregular distributed applications. Our approach is based on the actor/selector model, and  introduces the concept of a \textit{Partitioned Global Mailbox}. Subparts of the application that involve a large number of latency tolerant communication operations can use our system to achieve high throughput, while other parts of the application can keep using the PGAS communication system/runtime directly.  Actors are often used as the concurrency mechanism in newer languages such as Scala or Rust and in other domains such as the cloud. 
Thus Actor's addition to the PGAS ecosystem creates a low overhead path for users of other domains to develop high-performance computing applications without ramping up on non-blocking primitives. Moreover, we have shown that our Actor system beats the non-blocking operations in the state of the art communication libraries/systems by a handsome margin, thereby demonstrating the need to add/improve message aggregation in such libraries.
Our programming system also abstracts away low-level details of message aggregation (e.g.,  manipulating local buffers and managing progress and termination) so that the programmer can work with a high-level selector interface. 
Further, this approach can be integrated with the standard synchronization constructs in asynchronous task runtimes (e.g., async-finish, future constructs). 
Our Actor runtime is more than a message-aggregation system since it also supports user-defined active messages, which can support the migration of computation closer to data that is beneficial for irregular applications.
For the 2048 PE case,
our approach show a geometric mean performance improvement of 25.59$\times$ relative to the UPC versions, 19.83$\times$ relative to the OpenSHMEM  versions, and a geometric mean slowdown of only 1.09$\times$ relative to the Conveyors versions. 
These results suggest that the FA-BSP model offers a desirable point in the productivity-performance spectrum, with higher performance relative to PGAS models such as UPC and OpenSHMEM and higher productivity relative to the use of low-level hand-coded approaches for communication management and message aggregation.


In future work, we plan to support variable-sized messages since our current Mailbox implementation only accepts messages of a fixed size that is specified when creating a selector.  
Further, the original Selectors model~\cite{shams_selector} allows for operations such as mailbox priorities and enable/disable operations on mailboxes; support for such operations could enable richer forms of coordination logic across messages in our implementations.
Actors could also be used to implement non-blocking communication primitives thereby they can get the message aggregation capability without directly interfacing with low-level aggregation libraries. 
Finally, it would be interesting to explore compiler extensions to automatically translate from the natural version to our selector version, thereby directly improving the performance of natural PGAS programs.

